\documentclass[pdflatex,sn-mathphys-num]{sn-jnl}

\usepackage{graphicx}%
\usepackage{multirow}%
\usepackage{amsmath,amssymb,amsfonts}%
\usepackage{amsthm}%
\usepackage{mathrsfs}%
\usepackage[title]{appendix}%
\usepackage{xcolor}%
\usepackage{textcomp}%
\usepackage{manyfoot}%
\usepackage{booktabs}%
\usepackage{algorithm}%
\usepackage{algorithmicx}%
\usepackage{algpseudocode}%
\usepackage{listings}%

\theoremstyle{thmstyleone}%
\theoremstyle{thmstyletwo}%

\theoremstyle{thmstylethree}%

\begin{document}

\title[]{Quantum communication scheme to teleport arbitrary quantum state via discrete time quantum walks}


\author*[1]{\fnm{Rachana} \sur{Soni}}\email{E20SOE813@bennett.edu.in}

\author[2]{\fnm{Neelam} \sur{Choudhary}}\email{neelam.choudhary@bennett.edu.in}

\author[2]{\fnm{Navneet} \sur{Pratap Singh}}\email{navneet.singh@bennett.edu.in}

\affil*[1]{\orgdiv{School of Computer Science Engineering and Technology}, \orgname{Bennett University}, \orgaddress{\street{TechZone 2}, \city{Greater Noida}, \postcode{201310}, \state{U.P.}, \country{India}}}

\affil[2]{\orgdiv{School of Artificial Intelligence}, \orgname{Bennett University}, \orgaddress{\street{TechZone 2}, \city{Greater Noida}, \postcode{201310}, \state{U.P.}, \country{India}}}



\abstract{In this article, we propose a quantum communication protocol via 2-step discrete time quantum walks with two coins on a graph of 10 vertices containing both cycles and paths. Quantum walks are known for their ability to integrate quantum mechanics dynamics like superposition and entanglement during the procedure. We calculate the total final quantum state of the system as well as the recovery operators to rescue the initial quantum state back at the receiver's location. Our work provides a fundamental block for a quantum teleportation scheme on a specific series of cycle and path graphs.}


\maketitle
\section{Introduction}
{The theory of quantum teleportation was founded by Bennett et al.\cite{Bennett} in 1933, since then, it has become a field of extensive study  of quantum information theory for the researchers. It enables the transportation of an arbitrary quantum state between two parties. The quantum communication protocol includes quantum mechanics phenomena like superposition and entanglement. The fundamental protocol is result of study of entanglement and Bell states to project on quantum system on maximally entangled states. Quantum teleportation plays crucial role in quantum computation as measurement based quantum computation.Secure quantum teleportation can be utilized in quantum cryptography, as Quantum Key Distribution \cite{li2013discrete}. It expands the practical application of entanglement for transporting quantum information which has no classical counterpart and brings experimental realization of entanglement as a physical phenomenon.}

{ In the last decade, quantum walks applied as significant tool for transferring the quantum states in a designed network. Quantum walks have the ability to simulate quantum evolution and experiment the entanglement in physical aspect on graph based structures. These qualities make quantum walks a deserving candidate for quantum teleportation protocols. One can see extensive work related to DTQWs as important medium for state transfer and  develop algorithms in \cite{Amb}-\cite{Aharnov}, \cite{Shen}, \cite{kurzynski2011discrete}, \cite{yang2018quantum}. Multi-coin operators in DTQWs bring more complex and detailed insights for walk evolution as can be read in \cite{BTA}-\cite{ShangY}. Work related to theory of continuous time quantum walks can be found in \cite{christandl2004perfect}, \cite{mulken2007quantum}, \cite{mulken2011continuous}, \cite{RSoni1}, \cite{RSoni2}.}

{Generally, when we discuss about quantum teleportation, we name sender as Alice and the receiver as Bob, our goal to transfer unknown quantum state of Alice to Bob successfully. Quantum entanglement and measurement which are quantum mechanics events are utilized for this communication protocol. Classical communication is also used as encryptic code to make the communication secret and leak-proof. Hybrid mode makes the communication private and safer. In quantum walks, nodes work as qubits and walk evolution facilitate state transfer. The related teleportation work via quantum walks can be read in \cite{HQW0}-\cite{lovett2010universal}.
}

{ Key benefits of quantum walks as means of quantum teleportation are:
\begin{itemize}
    \item Graph symmetry and topology: Symmetric and topological qualities of graph structures provide controlled and smooth walk evolution. This may reduce errors and infidelity of state transfer \cite{aharonov1993quantum}, \cite{kempe2003quantum}.
    \item Scalability: Quantum walks can be scaled to larger family of graph of $n$ vertices, making them efficient for multi-qubit communication protocols \cite{rohde2011multi}.
    \item Degree of freedom as coin operator: The coin operators used in DTQWs provide degree of freedom to choose all the possible paths to quantum walker. Some popular choices for coin operators are Hadarmard, Grover and Fourier coin operators. Each has own special features and usability. 
    \item Entanglement distribution: In the procedure of applying evolution operator to initial state, the particles get entangled so there isn't separate need for entanglement generation among qubits. This is one of benefits of quantum teleportation via quantum walks which makes quantum walk ideal process for teleportation.
    \item Usage in quantum computing: Quantum walks can be incorporated into quantum algorithms for quantum state transfer of qubits and quantum computation on graph networks.
    
\end{itemize}}
{ Wang et al.\cite{Xue} established quantum teleportation via d-regular and d-complete graph and asked an open question that whether other graphs can be explored for the protocol they are proposing? Yang et al.\cite{YaCao} and Soni et al.\cite{RSoni3} provided teleportation work via quantum walks for arbitrary length on cycles, paths and hexagonal chain. We got motivated from the question in their article to find other graphs for successful teleportation, we here offer an irregular graph on 10 vertices which is joint of cycle and path alternatively. We are providing a fundamental block to work on arbitrary length of this specially designed graph. In our work, we formulate total final state of quantum walk evolution on cycled-path on 10 vertices after two steps of quantum walks and the recovery operators Bob the receiver can use to get the unknown quantum state back, making communication with high fidelity and least errors. Here, we used two coins discrete time quantum walks in two steps to achieve the goal.}

\section{Mathematical preliminaries} In this section, we introduce mathematical notations and framework used in procedure of two steps quantum walk with two coins for teleportation protocol. 

    \subsection{ Initial quantum state:} The initial quantum state of the system is defined by a ket notation in a \textit{Hilbert space} $\mathcal{H}$, $|\phi\rangle$ can be represented as a linear sum of orthonormal basis states $\{|j\rangle\}$:

\[
|\phi \rangle = \sum_{j}  a_j |j\rangle, \quad \text{with} \quad \sum_i |a_j|^2 = 1,
\]
where $a_j$ are complex amplitudes and $|j\rangle$ represents the basis ket vectors.
Total initial quantum state of the quantum system will be,
 $$ | \phi_0 \rangle = |1 \rangle \otimes \left( \sum_{k=0}^2 a_k |k \rangle) \right) \otimes |0\rangle $$ where first ket vector is related to position space(10 dimensional in our case) for particle $A_1$ and second and third ket vectors are related to coin spaces which will be $3$-dimensional spaces. 

\subsection{ Graph representation:}

The quantum walks operates on a graph structure. For our proposed  10-vertex graph, the position space is an \textit{10-dimensional Hilbert space} $\mathcal{H}_P$, spanned by $\{|0\rangle, |1\rangle, \dots, |9\rangle\}$. Coin space is spanned by maximum number of edge labeling. 

\subsection{Evolution operator of DTQWs} The quantum walk evolution is described by $U$ which is a unitary operator that consists of two operators- conditional shift and coin operators.
    \[
    U = S \cdot (C \otimes I),
    \]
\subsection{Conditional shift operator} The \textit{conditional shift operator} $S$ moves the quantum walker across all the possible paths of the graph. For a graph $G$, the operator $S$ acts as:

\[
S = \sum_{i \in V} |i + 1 \rangle \langle i| \otimes |k\rangle_c \langle k|,
\]
If there is an directed edge from $|i\rangle$ to $|i +1\rangle$ with edge label $k$.
\subsection{Coin operators} The \textit{coin operator} $C$ determines the internal dynamics as degree of freedom to choose the path for the walker. Here, in the two steps quantum walks, our first coin operator $C_A = I_3$ and second coin operator we choose Fourier coin $C_B= F_3$ for successful quantum teleportation of unknown quantum state so that no complex amplitude $a_j$ is lost during the process \\

    \textit{Fourier Coin} $F_3$ (for a 3-dimensional coin space):
    \[
    F_3 = \frac{1}{\sqrt{3}} \sum_{j=0}^2 \sum_{k=0}^2 e^{2\pi i jk / 3} |j\rangle \langle k|.
    \]

\subsection{Measurement basis for particles} After the two steps of quantum walk, Alice measures particle $A_1$ with basis $$\{|0 \rangle, |1 \rangle \ldots |9\rangle\} $$ and measurement basis for particle $A_2$ and $B$ will be $\{|f_0\rangle, |f_1\rangle, |f_2\rangle  \}$
where, $$|f_i \rangle =\frac{1}{\sqrt{3}} \sum_{k=0}^2 e^{2\pi i jk / 3} |k\rangle, \quad \text{for } j = 0, 1, 2.
 $$
 \subsection*{Fourier Basis States in Summation Form}

For a 3-dimensional Hilbert space, the Fourier basis states \( |f_j\rangle \) are defined as:

\[
|f_j\rangle = \frac{1}{\sqrt{3}} \sum_{k=0}^2 e^{2\pi i jk / 3} |k\rangle, \quad \text{for } j = 0, 1, 2.
\]

The detailed the Fourier basis states \( |f_0\rangle, |f_1\rangle, |f_2\rangle \) is as follows:

\begin{itemize}
    \item \textbf{\( |f_0\rangle \):}
    \[
    |f_0\rangle = \frac{1}{\sqrt{3}} \left( |0\rangle + |1\rangle + |2\rangle \right).
    \]

    \item \textbf{ \( |f_1\rangle \):}
    \[
    |f_1\rangle = \frac{1}{\sqrt{3}} \left( |0\rangle + e^{2\pi i / 3} |1\rangle + e^{4\pi i / 3} |2\rangle \right).
    \]

    \item \textbf{ \( |f_2\rangle \):}
    \[
    |f_2\rangle = \frac{1}{\sqrt{3}} \left( |0\rangle + e^{4\pi i / 3} |1\rangle + e^{8\pi i / 3} |2\rangle \right).
    \]
\end{itemize}

\subsection{Recovery operators} If the final quantum state collapses to some state $|\psi\rangle$ post-measurements of particles $A_1$ and $A_2$, then the \emph{recovery operator} $\mathcal{R}$ is applied to $|\psi\rangle$ to revise
the system back to its original unknown quantum state:
\[
\mathcal{R} \, |\psi\rangle \;=\; a_0|0\rangle \;+\; a_1|1\rangle \;+\; a_2|2\rangle.
\]

The coefficients $a_0, a_1, a_2$ are complex amplitudes satisfying 
$\lvert a_0\rvert^2 + \lvert a_1\rvert^2 + \lvert a_2\rvert^2 = 1$, 
representing the normalized superposition of the basis states 
$\{|0\rangle, |1\rangle, |2\rangle\}$ of the qutrit.

\section{Quantum teleportation scheme via $2-$steps DTQWs with two coins in cycled-path} 
The following graph is proposed for quantum communication scheme via two steps DTQWS with two coins. The boundary conditions, that is, how edges are directing with particular labels are expressed in Fig 1.\\

\begin{figure}[h!]
   \begin{center}
\includegraphics[width=9cm, height=4cm]{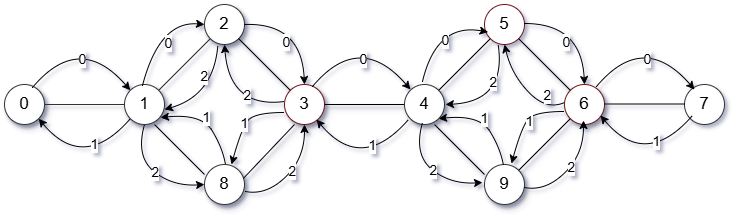}
\end{center} 
    \label{fig:fig1}
    \caption{Cycled-path consists of two $C_4$ and alternatively path on 10 vertices}
\end{figure}

\subsection{Quantum communication protocol scheme via DTQWs}
\begin{enumerate}
\item The goal is to send Alice's arbitrary quantum state $ \left( \sum_{k=0}^2 a_k |k \rangle \right) $ to Bob's location. 
    \item Alice has two particles $A_1$ and $A_2$. Bob has particle $B.$ At first step of quantum walk, particle $A_1$ and $A_2$ get entangled. Alice sends particle $A_2$ to Bob  and at the second step of quantum walk, particle $A_2$ and $B$ get entangled. Alice measures particle $A_2$ and tells the results to Bob.
    \item Alice measures final quantum state for particle $A_1$ with basis $ |0 \rangle, |1 \rangle \ldots |9\rangle $
    \item Alice measures particle $A_2$ with the following Fourier basis states,
  \begin{align*}      
&|f_0\rangle = \frac{1}{\sqrt{3}} \left( |0\rangle + |1\rangle + |2\rangle \right)\\
&|f_1\rangle = \frac{1}{\sqrt{3}} \left( |0\rangle + e^{2\pi i / 3} |1\rangle + e^{4\pi i / 3} |2\rangle \right)\\
&|f_2\rangle = \frac{1}{\sqrt{3}} \left( |0\rangle + e^{4\pi i / 3} |1\rangle + e^{8\pi i / 3} |2\rangle \right)
\end{align*}  
\item After the post-measurements, Alice tells her measurement results to Bob via classical communication. 
\item Bob reads the collapsed state and applies recovery operators to that collapsed state to get the initial quantum state of Alice back securely.
\end{enumerate}
Now we apply this scheme to proposed graph.
\begin{itemize}
\item The total initial quantum state of Alice is $$ | \phi_0 \rangle = |1 \rangle \otimes \left( \sum_{k=0}^2 a_k |k \rangle) \right) \otimes |0\rangle $$
\item {Conditional shift operator}
We calculated conditional shift operator for the proposed graph drawn in Fig 1. 
\begin{align*}
    S=  &\sum_{m=0}^7 |m+1 \rangle \langle m| \otimes |0\rangle \langle 0| + \sum_{m=1}^7 |m-1 \rangle \langle m| \otimes |1\rangle \langle 1| +\\ & |8 \rangle \langle 1| \otimes | 2 \rangle \langle 2| +|1 \rangle \langle 8| \otimes | 1 \rangle \langle 1| + |3 \rangle \langle 8| \otimes | 2 \rangle \langle 2| + \\ &|8 \rangle \langle 3| \otimes | 1 \rangle \langle 1| + |9 \rangle \langle 4| \otimes | 2 \rangle \langle 2| + |4 \rangle \langle 9| \otimes | 1 \rangle \langle 1| + \\ & |9 \rangle \langle 6| \otimes | 1 \rangle \langle 1| + |6 \rangle \langle 9| \otimes | 2 \rangle \langle 2|
\end{align*}
Rearranging the common terms
\begin{align*}
&S = \left( \sum_{m=0}^7 |m+1 \rangle \langle m| \right) \otimes |0\rangle \langle 0|  \\ & + \left(  \sum_{m=1}^7 |m-1 \rangle \langle i| + |1 \rangle \langle 8|+ |8 \rangle \langle 3| + |4 \rangle \langle 9| + |9 \rangle \langle 6| \right) \otimes |1\rangle \langle 1| \\ & + \Big(|8 \rangle \langle 1| + |3 \rangle \langle 8|+ |9 \rangle \langle 4| + |6 \rangle \langle 9| \\ &+ | 4 \rangle \langle 5| + |5 \rangle \langle 6| + |2 \rangle \langle 3| + |1 \rangle \langle 2 | \Big) \otimes | 2 \rangle \langle 2|
\end{align*}
\item The total final quantum state of the system 
$$ | \phi_t \rangle = U_2 U_1 | \phi_0 \rangle $$
 where, 
$$ U_1= S_1(I_{10} \otimes C_A \otimes I_2)$$
$$U_2 = S_2(I_{10} \otimes I_2 \otimes C_B ) $$
$$S_1 = S \otimes I_2$$
\begin{align*}
&S_2 = \left( \sum_{m=0}^7 |m+1 \rangle \langle m| \right)\otimes I_2 \otimes |0\rangle \langle 0|  \\ & + \left(  \sum_{m=1}^7 |m-1 \rangle \langle m| + |1 \rangle \langle 8|+ |8 \rangle \langle 3| + |4 \rangle \langle 9| + |9 \rangle \langle 6| \right) \\& \otimes I_2  \otimes |1\rangle \langle 1|  + \Big(|8 \rangle \langle 1| + |3 \rangle \langle 8|+ |9 \rangle \langle 4| + |6 \rangle \langle 9| \\ &+ | 4 \rangle \langle 5| + |5 \rangle \langle 6| + |2 \rangle \langle 3| + |1 \rangle \langle 2 | \Big) \otimes I_2 \otimes | 2 \rangle \langle 2|
\end{align*}
 $$C_A = I_3, C_B = F_3$$
  $$ F_3 = \frac{1}{\sqrt{3}} \sum_{j=0}^2 \sum_{k=0}^2 e^{2\pi i jk / 3} |j\rangle \langle k|.
 $$
\begin{align*}
&U_2 = \left( \sum_{m=0}^7 |m+1 \rangle \langle m| \right)\otimes I_2 \otimes |0\rangle \langle 0| F_3  \\ & + \left(  \sum_{m=1}^7 |m-1 \rangle \langle m| + |1 \rangle \langle 8|+ |8 \rangle \langle 3| + |4 \rangle \langle 9| + |9 \rangle \langle 6| \right) \\& \otimes I_2  \otimes |1\rangle \langle 1| F_3  + \Big(|8 \rangle \langle 1| + |3 \rangle \langle 8|+ |9 \rangle \langle 4| + |6 \rangle \langle 9| \\ &+ | 4 \rangle \langle 5| + |5 \rangle \langle 6| + |2 \rangle \langle 3| + |1 \rangle \langle 2 | \Big) \otimes I_2 \otimes | 2 \rangle \langle 2| F_3
\end{align*}
  \item After first step of quantum walk, let the quantum state will be $| \phi_t \rangle^1 $, then, 
  $$| \phi_t \rangle^1 = U_1 | \phi_0 \rangle $$
 
  or, 
 \begin{align*}
& | \phi_t \rangle^1= \Big[ \left( \sum_{m=0}^7 |m+1 \rangle \langle m| \right) \otimes |0\rangle \langle 0| \otimes I_2 \\ & + \left(  \sum_{m=1}^7 |m-1 \rangle \langle m| + |1 \rangle \langle 8|+ |8 \rangle \langle 3| + |4 \rangle \langle 9| + |9 \rangle \langle 6| \right) \\& \otimes |1\rangle \langle 1| \otimes I_2  + \Big(|8 \rangle \langle 1| + |3 \rangle \langle 8|+ |9 \rangle \langle 4| + |6 \rangle \langle 9| \\ &+ | 4 \rangle \langle 5| + |5 \rangle \langle 6| + |2 \rangle \langle 3| + |1 \rangle \langle 2 | \Big) \otimes | 2 \rangle \langle 2| \otimes I_2  \Big] \\& \Big[  |1 \rangle \otimes \left( \sum_{k=0}^2 a_k |k \rangle) \right) \otimes |0\rangle \Big]
\end{align*}
or 
\begin{align*}
   & | \phi_t \rangle^1= a_0 |200 \rangle + a_1 |010 \rangle + a_2 |820 \rangle
\end{align*}
\item Now, in the process of second step of quantum walk, operator $U_2$ will be applied to $| \phi_t \rangle^1$ to get total final quantum state of the system, and we get, 
\begin{align*}
   & | \phi_t \rangle= \dfrac{1}{\sqrt{3}}(a_0 |300 \rangle + a_0 |101 \rangle + a_1 |110\rangle + a_2|121 \rangle )
\end{align*}
\item Alice measures total final quantum state for the particle $A_1$ with basis $|0 \rangle, |1 \rangle, \ldots, |9\rangle $ and if she gets $|1 \rangle$, the state collapse to \begin{equation}
\dfrac{1}{\sqrt{3}}(a_0 |01 \rangle + a_1 |10\rangle + a_2|21 \rangle )
\end{equation}.  
\item Alice measures state(1) for the particle $A_2$ with basis $|f_0\rangle,|f_1\rangle,|f_2\rangle $ and tells her measurement results to Bob.\\

Expressing state(1) in terms of\( |f_0\rangle, |f_1\rangle, |f_2\rangle \) and collecting their coefficients,

\begin{enumerate}
    \item \textit{Coefficient of \( |f_0\rangle \):}
    \[
    a_0 |1\rangle + a_1 |0\rangle + a_2 |1\rangle.
    \]

    \item \textit{Coefficient of \( |f_1\rangle \):}
    \[
    a_0 |1\rangle + a_1 e^{2\pi i / 3} |0\rangle + a_2 e^{4\pi i / 3} |1\rangle.
    \]

    \item \textit{Coefficient of \( |f_2\rangle \):}
    \[
    a_0 |1\rangle + a_1 e^{4\pi i / 3} |0\rangle + a_2 e^{8\pi i / 3} |1\rangle.
    \]
\end{enumerate}

 To recover initial quantum state of Alice back, Bob applies recovery operators to coallapsed states that is coefficients of $|f_i\rangle, i=0,1,2$, accordingly, for every possible measurements, as given in the below table.
\end{itemize}
\begin{table}[h]
\caption{\label{measurementresults}Measurement and Recovery Operators.}
\begin{tabular}{@{}lll}
\hline
\textbf{Measurement \( A_1 \)}&\textbf{Measurement \( A_2 \)}&\textbf{Recovery Operators}\\
\hline
\(|1 \rangle\)&\( |f_0 \rangle \) &\(|0\rangle \langle 1 | + |1 \rangle \langle 0| + |2 \rangle \langle 1| \) \\ 
\(|1 \rangle\)&\( |f_1 \rangle \)&\((|0\rangle \langle 1 | + |1 \rangle \langle 0| + |2 \rangle \langle 1|) P_1 \)\\  
\(|1 \rangle\) & \( |f_2 \rangle \) & \((|0\rangle \langle 1 | + |1 \rangle \langle 0| + |2 \rangle \langle 1|) P_2 \) \\
\hline
\end{tabular}

\end{table}
Here $P_1$ and $P_2$ are phase correction operators, \\
$$P_1=e^{-2\pi i /3}|0 \rangle \langle 0|+e^{-4 \pi i /3}|1 \rangle \langle 1| + |2\rangle \langle 2|$$ and
$$P_2= e^{-4\pi i /3}|0 \rangle \langle 0|+e^{-8 \pi i /3}|1 \rangle \langle 1| + |2\rangle \langle 2|$$
\section{Conclusion and future scope}
{In this research work, we study the quantum teleportation protocol scheme via DTQWs on a graph of 10 vertices which consists of  path and cycle alternatively. The symmetry and topology of the proposed graph allows successful teleportation of arbitrary quantum state from the sender Alice to the receiver Bob's end. Yang et al.\cite{YaCao}
published quantum teleportation results for N-cycle and N-path graph. We joined both kind of graph in a alternate way to create a new graph structure as given in Fig. 1, to examine teleportation via quantum walk. We formulate total final quantum state after two steps of quantum walks and recovery operator to get complete fidelity of state transfer.}

{Our work opens up new roads for next step of research where the final quantum state of the system and recovery operator can be achieved for arbitrary length of such graph. We examined two steps quantum walks for the communication, one can explore many steps and many coin version of this communication scheme. }


\section{Data Availability Statements}

Data sharing is not applicable to this research article as no dataset is generated or analyzed during the study.

\section{Conflict of interest statement}
On behalf of all authors, the corresponding author states that there is no conflict of interest.

\end{document}